\documentclass[letterpaper]{article} 
\usepackage{aaai25}  
\usepackage{times}  
\usepackage{helvet}  
\usepackage{courier}  
\usepackage[hyphens]{url}  
\usepackage{multirow}
\usepackage{booktabs}
\usepackage{textcomp}
\usepackage{array}
\usepackage{subcaption}
\usepackage{graphicx} 
\usepackage{natbib}  
\usepackage{caption} 
\usepackage{algorithm}
\usepackage{algorithmic}

\usepackage{newfloat}
\usepackage{listings}
\DeclareCaptionStyle{ruled}{labelfont=normalfont,labelsep=colon,strut=off} 
\floatstyle{ruled}
\newfloat{listing}{tb}{lst}{}
\floatname{listing}{Listing}
\usepackage{amssymb}
\usepackage{amsmath}

\title{DisCo: Graph-Based Disentangled Contrastive Learning for Cold-Start \\ Cross-Domain Recommendation}
\def\method{DisCo}
\author{
    Hourun Li\equalcontrib\textsuperscript{\rm 1,2},
    Yifan Wang\equalcontrib\textsuperscript{\rm 3},
    Zhiping Xiao\textsuperscript{\rm 4}\thanks{Corresponding authors.},
    Jia Yang\textsuperscript{\rm 2},
    Changling Zhou\textsuperscript{\rm 2},\\
    Ming Zhang\textsuperscript{\rm 1}\textsuperscript{†},
    Wei Ju\textsuperscript{\rm 5}\textsuperscript{†}
}
\affiliations{
    \textsuperscript{\rm 1}  State Key Laboratory for Multimedia Information Processing, School of Computer Science, \\PKU-Anker LLM Lab, Peking University, Beijing, China \\
    \textsuperscript{\rm 2} Computer Center, Peking University, Beijing, China \\
    \textsuperscript{\rm 3} School of Information Technology \& Management, University of International Business and Economics, Beijing, China\\
    \textsuperscript{\rm 4} Paul G. Allen School of Computer Science and Engineering, University of Washington, Seattle, WA, USA \\
    \textsuperscript{\rm 5} College of Computer Science, Sichuan University, Chengdu, China\\
    lihourun@stu.pku.edu.cn, \{yangj, zclfly, mzhang\_cs\}@pku.edu.cn,\\
    yifanwang@uibe.edu.cn, patxiao@uw.edu, juwei@scu.edu.cn
}

\usepackage{bibentry}

\begin{document}
\maketitle
\begin{abstract}
Recommender systems are widely used in various real-world applications, but they often encounter the persistent challenge of the user cold-start problem. 
Cross-domain recommendation (CDR), which leverages user interactions from one domain to improve prediction performance in another, has emerged as a promising solution. 
However, users with similar preferences in the source domain may exhibit different interests in the target domain. 
Therefore, directly transferring embeddings may introduce irrelevant source-domain collaborative information. 
In this paper, we propose a novel graph-based disentangled contrastive learning framework to capture fine-grained user intent and filter out irrelevant collaborative information, thereby avoiding negative transfer. 
Specifically, for each domain, we use a multi-channel graph encoder to capture diverse user intents. 
We then construct the affinity graph in the embedding space and perform multi-step random walks to capture high-order user similarity relationships. 
Treating one domain as the target, we propose a disentangled intent-wise contrastive learning approach, guided by user similarity, to refine the bridging of user intents across domains.
Extensive experiments on four benchmark CDR datasets demonstrate that \method{} consistently outperforms existing state-of-the-art baselines, thereby validating the effectiveness of both \method{} and its components.
\end{abstract}

%

\section{Introduction}
\begin{figure}[t]
\centering
\includegraphics[width=1\linewidth]{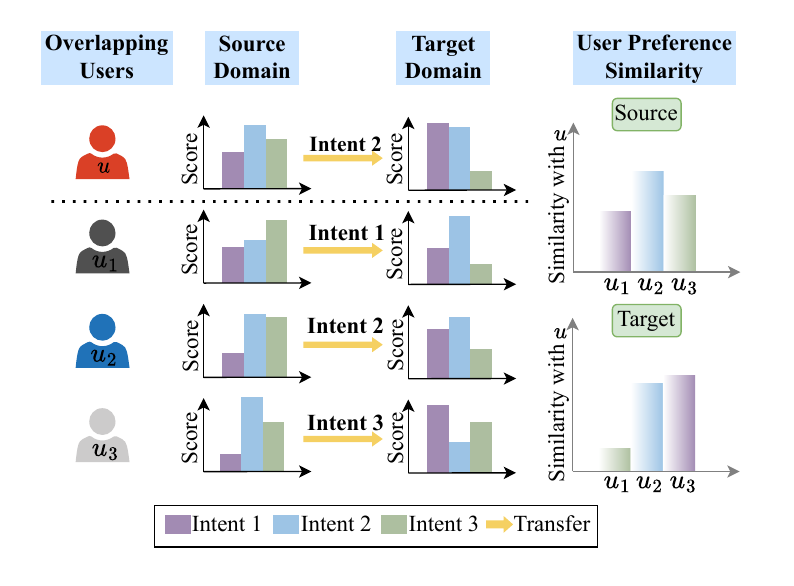}
\vspace{-0.5cm}
\caption{An illustration of user similarity distortion in CDR. Users may share diverse intents across domains. As a result, for the target user $u$, some users (i.e., $u_1$, $u_3$) may have similar preferences in the source domain but differ significantly in the target domain, leading to negative transfer.}
\label{fig:1}
\vspace{-0.5cm}
\end{figure}


The rapid expansion of Internet services has deeply integrated online platforms into our daily lives, resulting in an enormous increase in digital information~\cite{zhu2016fuzzy,zhang2020str,wang2022he,wang2022ad,ju2022kernel,qu2024graph,ju2024hypergraph}. 
Consequently, personalized recommender systems are crucial for guiding users through extensive options to identify items that align with their preferences. Collaborative Filtering (CF), which models relationships between users and dependencies among items, has seen significant success in recommendation systems. 
However, these CF-based methods persistently suffer from the cold-start problem, 
where new users lack sufficient observed interactions, making it challenging to learn effective representations for recommendations.

To alleviate the above issue, cross-domain recommendation (CDR) has garnered considerable attention in recent years. 
The main idea of CDR methods is to leverage user interaction data from related domains to improve prediction accuracy in the target domain. 
A particularly challenging aspect is cold-start CDR, where users have interactions with items in one domain but not in the other. Traditional Embedding and Mapping (EMCDR) paradigms~\cite{man2017cross,kang2019semi,salah2021towards} encode user preferences in two domains separately and learn a mapping function from the source to the target domain, which often overlooks the diverse user-specific preferences~\cite{wang2022deep}. 
Additionally, some meta-learning approaches~\cite{zhu2021transfer,zhu2022personalized,guan2022cross,li2024cdrnp} treat different user CDR as individual tasks to achieve user-specific preference transfer, where the transformed preference can be utilized as the initial embedding for the cold-start users.

Nevertheless, it still remains challenging to address the negative transfer issue for CDR~\cite{li2024aiming}. 
Since information from the source domain is not always relevant to the target domain, indiscriminately incorporating source domain data during training may lead to negative effects. 
Recent efforts attempt to address this problem by designing disentangled user representations that transfer only the relevant information. 
For example, CDRIB~\cite{cao2022cross} introduces two information bottleneck regularizers to encourage the learned representation to encode domain-shared information while limiting domain-specific information.
DisenCDR~\cite{cao2022disencdr} learns disentangled representations to separate domain-shared and domain-specific user preferences. 
UniCDR~\cite{cao2023towards} proposes a unified framework to capture domain-shared and domain-specific user preferences for different CDR scenarios. 

Despite their promising performance, we find that most of the cold-start CDR methods focus on the user-embedding transfer, neglecting the more detailed underlying intent and preference correlations among users. 
In fact, the complex formation of user preferences requires to infer the latent intent under implicit interaction. 
Meanwhile, users who exhibit similar preferences in the source domain may have different interests in the target domain. 
Therefore, collaborative knowledge from the source domain can become irrelevant or even noisy in the target domain, ending up counter-effective.
We demonstrate such examples in Figure ~\ref{fig:1}, where user preferences are fine-grained with individual intents. 
Suppose we aim to learn the user preference of $u$ in the target domain. 
Given the preferences of $u_1$, the similarity in the source domain may not hold in the target domain. 
As a result, learning from the irrelevant preferences of these known users can introduce bias and induce sub-optimal performance.






Motivated by the above observation, we propose \textbf{DisCo}, a novel graph-based \underline{Dis}entangled \underline{Co}ntrastive learning framework for cold-start CDR, which is capable of capturing multiple intents of users and filtering our irrelevant source domain collaborative information. 
Specifically, we first utilize a multi-channel graph encoder to discern the underlying intent of users. 
Then, given the affinity graph calculated in the embedding space of two domains, we perform the multi-step random walks for each anchor user to obtain the high-order user similarities of each domain. 
Besides, we treat one of the domains as the target domain and propose an intent-wise contrastive architecture that performs both intra-domain and inter-domain contrastive learning to retain user similarity information for the target domain with the help of a cross-domain decoder. 
In this way, when introducing the source domain information, target domain-specific user preferences could be preserved while irrelevant collaborative information could be effectively filtered out by regularizing the rationale that bridging two domains.

In this paper, we make the following contributions:
\begin{itemize}
    \item \textit{Conceptual:} We highlight the negative transfer problem in cold-start CDR and propose to learn the more detailed disentangled user intent representation enhanced by the interaction graph to filter out irrelevant information. 
    \item \textit{Methodological:} We propose a novel intent-wise contrastive learning framework to retrain the user similarity information of the target domain, which could preserve the target domain-specific user preference and explicitly exploit the rationale that bridging two domains.
    \item \textit{Experimental:} We conduct extensive experiments on various public datasets to evaluate \method{}. Experimental results demonstrate the superiority of our proposed framework for the cold-start CDR task. The code is released on \url{https://github.com/HourunLi/2025-AAAI-DisCo}
\end{itemize}

\section{Related Work}
\subsection{Cross-Domain Recommendation}


Recent CDR efforts can be categorized into two types. 
Intra-domain CDRs~\cite{singh2008relational, li2009can,hu2018conet,li2020ddtcdr,liu2020cross}  address data sparsity by transferring abundant information from other domains for domains with limited user interactions. 
CBT~\cite{li2009can} introduces a cluster-level pattern matrix to transfer the rating patterns. 
CMF~\cite{singh2008relational} adapts the matrix decomposition to jointly factorize rating matrices across domains and shares the user latent factors. CoNet~\cite{hu2018conet}, DDTCDR~\cite{li2020ddtcdr}, and Bi-TGCF~\cite{liu2020cross} leverage deep models with information transfer modules. 
In contrast, inter-domain CDRs, or cold-start CDRs, tackle the more challenging task of recommending items to cold-start (non-overlapping) users with no prior interactions in a target domain.
EMCDR~\cite{man2017cross} pre-trains user embeddings of each domain and maps them by overlapping users. 
SSCDR~\cite{kang2019semi} and SA-VAE~\cite{salah2021towards} extend mapping with semi-supervised learning and variational autoencoder, respectively. 
PTUPCDR~\cite{zhu2022personalized} and TMCDR~\cite{zhu2021transfer} treat different user CDRs as individual tasks and use personalized meta-networks for user-specific preference transfer. 
UniCDR~\cite{cao2023towards} unifies intra- and inter-CDR in a single framework. 
However, most methods focus on user embedding transfer, neglecting underlying intent and the domains-specific user preferences.   

\subsection{Disentangled Representation Learning}
Disentangled representation learning seeks to develop factorized representations that can effectively distinguish and separate the underlying explanatory factors within observed data~\cite{bengio2013representation}. 
Existing researches have primarily focused on computer vision~\cite{higgins2016beta, chen2016infogan}, natural language processing~\cite{cheng2020improving, wang2022disencite}, and graph learning~\cite{ma2019disentangled,li2021disentangled,wang2020disenhan,wang2024disensemi,ju2024surveyb}.
For example, InfoGAN~\cite{chen2016infogan} separates representation into noise and an additional class code, estimating mutual information (MI) between the class code and corresponding data for controllable image generation. 
DGCL~\cite{li2021disentangled} introduces a self-supervised factor-wise contrastive learning framework to learn disentangled graph representation. 
Recently, there has been a notable surge of interest in applying disentangled representation learning techniques to recommendation systems~\cite{ma2019learning,wang2020disentangled,wang2022disenctr,qin2023disenpoi}. 
MacridVAE~\cite{ma2019learning} analyzes user behavior data via disentangling representations into macro and micro levels to investigate hierarchical user intentions. DisenHAN~\cite{wang2020disentangled} iteratively identifies the dominant aspects of various relations within a Heterogeneous Information Network (HIN) for the recommendation. 
For CDR task, DR-MTCDR~\cite{guo2023disentangled}, CDRIB~\cite{cao2022cross}, DisenCDR~\cite{cao2022disencdr} and GDCCDR~\cite{liu2024graph} separate user preferences into domain-shared and domain-specific parts, focusing on identifying and transferring the shared aspects to improve performance. 
In this paper, we propose a disentangled learning framework to capture more detailed user intent and filter out irrelevant collaborative information to avoid negative transfer. 

\section{Notations and Problem Definition}
\subsubsection{Notations.} Given the two domains $\mathcal{S}$ and $\mathcal{T}$, let $\mathcal{D}_\mathcal{S}=\{\mathcal{U}_\mathcal{S},\mathcal{V}_\mathcal{S}, \mathcal{E}_\mathcal{S}\}$ and $\mathcal{D}_\mathcal{T}=\{\mathcal{U}_\mathcal{T},\mathcal{V}_\mathcal{T},\mathcal{E}_\mathcal{T}\}$ denote the corresponding data of source and target domains, where $\mathcal{U}$, $\mathcal{V}$ and $\mathcal{E}$ denote the user, item and interaction set in the domain. The binary interaction matrix can be represented as $Y=\{y_{uv}\}\in\mathbb{R}^{|\mathcal{U}|\times|\mathcal{V}|}$, where $y_{uv}=1$ if $(u,v)\in\mathcal{E}$, otherwise, $y_{uv}=0$. 
Given the interaction matrix of source domain $Y_\mathcal{S}$ and target domain $Y_\mathcal{T}$, we construct the bipartite graphs $\mathcal{G}_\mathcal{S}$ and $\mathcal{G}_\mathcal{T}$ to depict the user-item relations of two domains. 
In particular, the user sets $\mathcal{U}_{\mathcal{S}}$ and $\mathcal{U}_{\mathcal{T}}$ contain an overlapping user subset $\mathcal{U}_o$. Then, the user set can be formulated as $\mathcal{U}_{\mathcal{S}}=\{\mathcal{U}_s,\mathcal{U}_o\}$, $\mathcal{U}_{\mathcal{T}}=\{\mathcal{U}_t,\mathcal{U}_o\}$, where $\mathcal{U}_s$ and $\mathcal{U}_t$ are the non-overlapping user set in each domain.

\subsubsection{Problem Definition.} Given the observed data $\mathcal{D}_\mathcal{S}$ and $\mathcal{D}_\mathcal{T}$ from the source and target domains, cold-start CDR aims to make recommendations in the target domain for non-overlapping users who are only observed in the source domain. 
Formally, given a user $u \in \mathcal{U}_s$ from the source domain, we seek to learn disentangled representations of user $u$ that capture diverse intents, in order to learn the prediction function $\hat{y}_{uv}=\mathcal{F}(u,v|\Theta, \mathcal{G}_\mathcal{S},\mathcal{G}_\mathcal{T})$, where item $v\in \mathcal{V}_\mathcal{T}$ is from the target domain. 
We use $\hat{y}_{uv}$ to denote the probability of user $u$ engaging with item $v$ and use $\Theta$ to denote the set of model parameters for the prediction function $\mathcal{F}$.


%
\section{The Proposed Framework}
\begin{figure*}[t]
    \centering
    \includegraphics[width=0.95\linewidth]{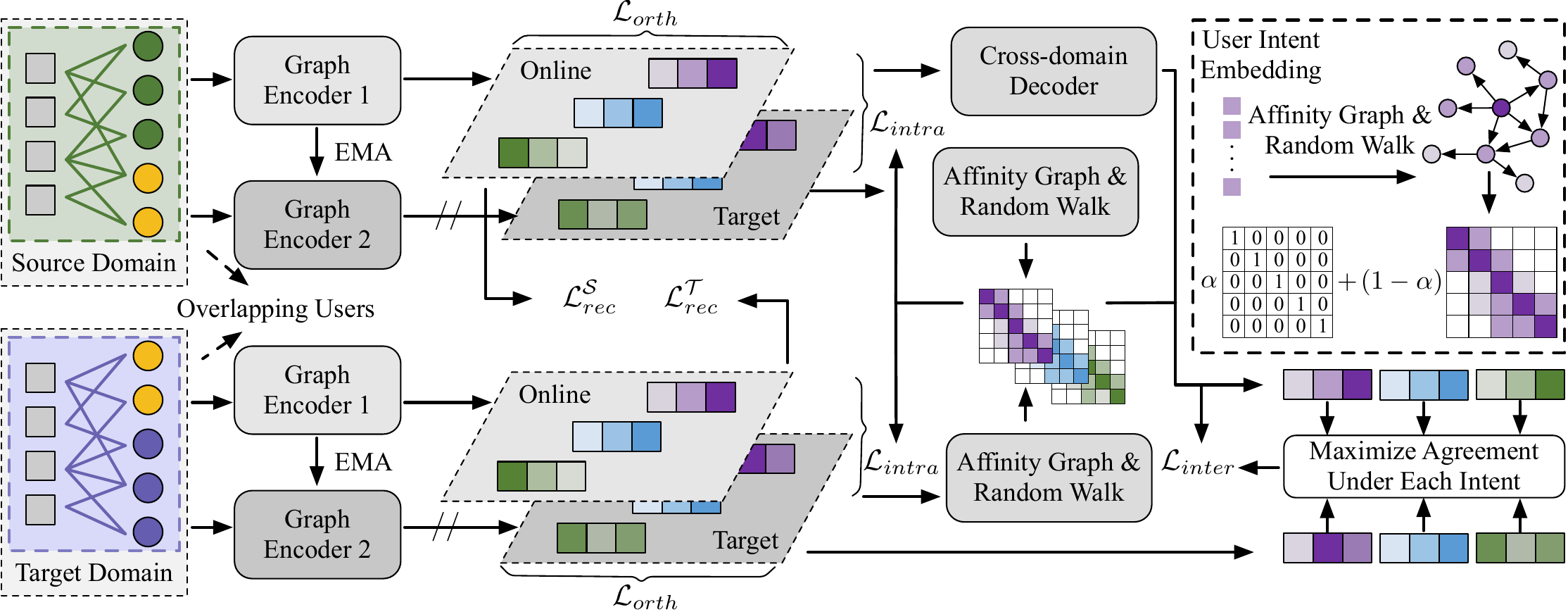}
    \caption{Illustration of the proposed framework \method{}.}
    \label{fig:2}
\end{figure*}
\subsection{Overview}
The fundamental concept of our framework is to alleviate the negative transfer problem when introducing the source domain information for cold-start CDR. 
As shown in Figure~\ref{fig:2}, there are three components in our framework. 
Given the interaction data $\mathcal{D}_\mathcal{S}$ and $\mathcal{D}_\mathcal{T}$ from the source and target domains, we construct bipartite graph $\mathcal{G}_\mathcal{S}$ and $\mathcal{G}_\mathcal{T}$ and extract fine-grained user intents for each domain via the proposed graph encoder. 
Next, affinity graphs are constructed within the corresponding embedding spaces of both domains.
By applying multi-step random walks based on affinity graphs, we progressively obtain user similarities as the collaborative information in both the source and target domain. 
And a disentangled intent-wise contrastive learning framework is proposed to retain the user similarity relationships while explicitly identifying the rationale between two domains to avoid the negative transfer. 
Finally, the disentangled user intents of the cold-start user are used to predict matching scores for cross-domain recommendations.
\subsection{Disentangled User Intent Graph Encoder}
As graph neural networks (GNNs) show a strong ability to model user interactions for the recommender~\cite{ju2024towards,ju2024survey}, we leverage GNNs as the backbone to capture the fine-grained user intents in each domain. 
For the constructed graph from the source and target domain, denoted as $\mathcal{G}_\mathcal{S}$ and $\mathcal{G}_\mathcal{T}$, we iteratively update the representations of user and item nodes through a message-passing mechanism. 
Taking user $u$ as an example, the $\text{GNN}^l(\cdot)$ can be:
\begin{equation}
\begin{split}
    z_u^{l+1} &= \sigma\Bigl(W_1^l z_u^l+ \sum_{v\in\mathcal{N}_u}\frac{1}{\sqrt{|\mathcal{N}_u||\mathcal{N}_v|}}m_{uv}\Bigr),\\
    m_{uv} &= W_1^lz_v^l + W_2^l(z_v^l\odot z_u^l),
\end{split}
\end{equation}
where $z_u^l$ denotes the updated user embedding at $l$-th layer, $m_{uv}$ represents the neighbor messages from the interaction item set $\mathcal{N}_u$, $\sigma(\cdot)$ is the activate function (i.e., LeakyReLU) and $W_{1/2}$ are trainable parameters.
After applying $L$ traditional message-passing layers, we obtain the extracted embedding matrix $Z^L$. 
We further utilize a graph-disentangled layer to extract the more detailed user intent with a separate channel. 
For each channel, we adopt $\text{GNN}_k^{L+1}$ and the disentangled user/item representation can be obtained as: 
\begin{equation}
    Z_k = \text{GNN}_k^{L+1}(Z^L, Y),
\end{equation}
where $k\in[1,K]$ and $K$ represents the number of intents.

\subsection{Disentangled Intent-wise Contrastive Learning}



To tackle the differences in collaborative information between domains, we propose a disentangled intent-wise contrastive learning framework that uses user similarity to capture fine-grained user intents and transfer relevant information from the source domain to the target domain.

\subsubsection{Intra-Domain Contrast.}
For each domain, we introduce two Siamese encoders, an online and a momentum-based target encoder, to generate effective self-supervised signals for fine-grained user intent learning. 
Specifically, both the online and target encoders share the same architecture and process the interaction data of each domain to obtain the user intents, denoted as  $Z_k$ and $\hat{Z}_k$.
The target encoder is updated using the Exponential Moving Average (EMA) of the online encoder. 
Then, the user similarity in a domain for the batch instances under intent $k$ can be defined as:
\begin{equation}
    R_{ij} = \exp(-\|\hat{z}_{i,k}-\hat{z}_{j,k}\|/\tau),
\end{equation}
where $\hat{z}_{i,k}$ represents the $k$-th channel of the output target embedding for user $u_i\in\mathcal{U}$. 
By treating user similarity as the edge weight between users, a fully connected user affinity graph $R$ could be constructed. 
And we normalize it in a row-wise manner to obtain the random walk transition matrix $\tilde{R}$. 
Considering a $d$ step random walks~\cite{lu2024decoupled}, the high-order user similarity under intent $k$ could be:
\begin{equation}
\label{randomWalk}
    T_k = \alpha I + (1-\alpha)\tilde{R}^{d},
\end{equation}
where $\alpha$ is a trade-off parameter between self and user pair similarities. We take the similarity as the pseudo target and formalize the intra-domain contrastive loss as:
\begin{equation}
    \mathcal{L}_{intra} = \sum_{u=1}^{|\mathcal{U}|}\sum_{k=1}^K H(T_k,\rho(Z_k,\hat{Z}_k)),
\end{equation}
where $H(\cdot,\cdot)$ denotes the cross entropy, $\rho(\cdot,\cdot)$ is the pairwise similarity with the row-wise normalization manner:
\begin{equation}
    [\rho(Z_k, \hat{Z}_k)]_{ij}=\frac{\exp(\phi(z_{i,k},\hat{z}_{j,k})/\tau)}{\sum_{j'=1}^{\mathcal{|\mathcal{U}|}}\exp(\phi(z_{i,k},\hat{z}_{j',k})/\tau)},
\end{equation}
where $\tau$ is the temperature parameter, and $\phi(\cdot,\cdot)$ denotes the similarity function. 
Moreover, we aim to learn disentangled user representation for diverse user intents, which means $z_{i,k}$ and $z_{i,k'}$ under channel $k$ and $k'$ are independent of each other. 
Since orthogonality is a specific instance of linear independence, we loose the representation constraint to orthogonality (as Equation~\ref{orth}), a method has been demonstrated to be effective in several previous studies ~\cite{liang2020attributed,wang2024disensemi}.
\begin{equation}
\label{orth}
   \mathcal{L}_{orth} = |{Z^{\mathrm{T}}_k}Z_k-I| + |{\hat{Z}^{\mathrm{T}}_k}\hat{Z}_k-I|.
\end{equation}




\subsubsection{Inter-Domain Contrast.} 
Given the extracted fine-grained user intents, we aim to transfer cross-domain information 
while maintaining distinct user collaborative information for each domain to highlight domain differences. 
Thus, simply maximizing the consistency between domains may lead to a loss of domain-specific information. 
To solve this issue, we introduce a user representation decoder that could preserve the domain-specific information while capturing the domain-shared information between the source and target domains. 
To be specific, for an overlapping user $u_i\in\mathcal{U}_o$, we project the source domain embedding $z_{i,k}^s$ to learn its corresponding target domain embedding.
\begin{equation}
    e_{i,k}^{(s\rightarrow t)} = g^{(s\rightarrow t)}(z_{i,k}^s),
\end{equation}
where $g(\cdot)$ denotes the decoder, and we simply implemented it with an MLP. 
To maintain user collaborative information in the target domain and prevent it from being disrupted by the incoming source domain information, we compute the similarity among overlapping users in the target domain by integrating all intents, namely, $T^t=1/K\sum_{k=1}^KT_k^t$. 
In this way, we use it as a supervision signal and the inter-domain contrastive loss can be defined as:
\begin{equation}
   \!\!\mathcal{L}_{inter} \!=\!\! \sum_{u_i=1}^{|\mathcal{U}_o|}\!H(T^t, p(u_j|u_i))\!=\!-\!\!\sum_{u_i=1}^{|\mathcal{U}_o|}\!T^t_{ij}\log p(u_j|u_i)\!
\end{equation}
where $p(u_j|u_i)$ denotes the user similarity identification task. 
Considering the latent fine-grained user intents, the cross entropy objective can be defined as:
\begin{equation}
    H(T^t, p(u_j|u_i))=-T_{ij}^t\log \mathbb{E}_{p(k|u_i)}p(u_j|u_i,k),
\end{equation}
where $p(k|u_i)$ is the prior distribution over latent intents for user $u_i$. 
We introduce $K$ intent prototypes $\{c_k\}_{k=1}^k$, and the prior distribution can be obtained as:
\begin{equation}
    p(k|u_i) = \frac{\exp(\phi(e^{s\rightarrow t}_{i,k},c_k))}{\sum_{k'=1}^K\exp(\phi(e^{s\rightarrow t}_{i,k'},c_{k'}))}.
\end{equation}
$p(u_j|u_i,k)$ denotes the user similarity under intent $k$. 
However, directly minimizing the objective is difficult because of latent user intents. Instead, we adopt an EM algorithm to solve the problem, where the posterior distribution $p(k|u_i,u_j)$ is defined with Bayes' theorem:
\begin{equation}
    p(k|u_i,u_j)=\frac{p(k|u_i)p(u_j|u_i,k)}{\sum_{k=1}^K p(k|u_i)p(u_j|u_i,k)}.
\end{equation}
$p(k|u_i,u_j)$ reflects how well the $k$-th intent aligns $u_i$ and $u_j$ in source and target domains. However, we cannot compute the posterior distribution because calculating $p(u_j|u_i,k)$ would require considering all users in the dataset. Therefore, we maximize the evidence lower bound (ELBO) of the log-likelihood, which can be defined as:
\begin{equation}
\begin{split}
    \log p(u_j|u_i)\geq&\mathbb{E}_{q(k|u_j,u_i)}[\log p(u_j|u_i,k)]\\
    &-D_{KL}(q(k|u_j,u_i)\|p(k|u_i)),
\end{split}
\end{equation}
where $q(k|u_j,u_i)$ is a variational distribution used to approximate the posterior distribution, calculated as follows:
\begin{equation}
    q(k|u_j,u_i)=\frac{p(k|u_i)\hat{p}(u_j|u_i,k)}{\sum_{k=1}^K p(k|u_i)\hat{p}(u_j|u_i,k)},
\end{equation}
where $\hat{p}(u_j|u_i,k)$ is the user similarity identification under intent $k$, calculated within a mini-batch $\mathcal{B}\subset\mathcal{U}_o$:
\begin{equation}
    \hat{p}(u_j|u_i,k)=\frac{\exp(\phi(e_{i,k}^{s\rightarrow t},\hat{z}_{j,k}^t))}{\sum_{j'\in\mathcal{B}}^{|\mathcal{B}|}\exp(\phi(e_{i,k}^{s\rightarrow t},\hat{z}_{j',k}^t))}.
\end{equation}
Notice the optimizing process is a variation of Variational EM algorithm, where we infer $q(k|u_j,u_i)$ at the E-step and optimize the ELBO at the M-step.




%

\subsection{User Intent Adaptation and Prediction}
For any single domain, we obtain the disentangled item representation $z_v=\{z_{v,1},\dots,z_{v,K}\}$ and the disentangled user representation $z_u=\{z_{u,1},\dots,z_{u,K}\}$, which represent the diverse $K$ user intents. 
We use the inner product of user and item representations for each intent as the predictive function to estimate the likelihood of their interaction under that specific intent. 
Thereafter, we weight the sum of all intents to obtain the overall user preference,
\begin{equation}
    r_{uv}=\sum_{k=1}^K p(k|u)\cdot z_{u,k}^{\mathrm{T}}z_{v,k}.
\end{equation}
We constrain the user preference score in the range of $[0,1]$ as the final matching score, which can be computed using a probabilistic function like the logistic function:
\begin{equation}
    \hat{y}_{uv} = \text{sigmoid}(r_{uv})=\frac{1}{1+\exp(r_{uv})}.
\end{equation}
Following the previous work~\cite{cao2023towards}, we use the binary cross entropy (BCE) loss and train the model using a negative sampling strategy. 
Taking the target domain $\mathcal{T}$ as an example, the recommendation loss is defined as follows:
\begin{equation}
    \!\mathcal{L}_{rec}^{\mathcal{T}} = -\!\sum_{(u,v)\in\mathcal{D}_{\mathcal{T}}}\log\hat{y}_{uv}-\!\sum_{(u,v^-)\in\mathcal{D}_{\mathcal{T}}^-}\log(1-\hat{y}_{uv^-}),
\end{equation}
where $\mathcal{D}_{\mathcal{T}}^-$ denotes the negative pairs uniformly sampled from unobserved interactions in the target domain.

\subsection{Objective Function}
The final loss term consists of two parts: 1) the contrastive loss, including the intra-domain contrastive loss $\mathcal{L}_{intra}$, the orthogonality loss $\mathcal{L}_{orth}$ and the inter-domain contrastive loss $\mathcal{L}_{inter}$.
2) the source and target domain prediction loss $\mathcal{L}_{rec}^{\mathcal{S}}$ and $\mathcal{L}_{rec}^{\mathcal{T}}$. And the overall objective is defined as:
\begin{equation}
\label{objectiveFunc}
\begin{split}
    &\mathcal{L}_{rec} = \mathcal{L}_{rec}^{\mathcal{S}} + \mathcal{L}_{rec}^{\mathcal{T}},\\
    &\mathcal{L}_{contra} = \beta\mathcal{L}_{inter} + (1-\beta)(\mathcal{L}_{intra} + \gamma\mathcal{L}_{orth}),\\
    &\mathcal{L}_{total} = \lambda\mathcal{L}_{contra}+(1-\lambda)\mathcal{L}_{rec}.
\end{split}
\end{equation}
where $\beta$, $\gamma$ and $\lambda$ are hyper-parameters that control the weights of each part of losses. We optimize the overall objective in a mini-batches manner.

\section{Experiment}
\begin{table*}[]
\centering
\setlength{\tabcolsep}{2.5pt}
\resizebox{\textwidth}{!}{%
\begin{tabular}{c|c|ccc|cccccc|c}
\hline\hline
\multirow{2}{*}{Dataset} & \multirow{2}{*}{Metric@10} & \multicolumn{3}{c|}{Single-Domain Methods} & \multicolumn{6}{c|}{Cross-domain Methods} & Ours \\ \cline{3-12} 
 &  & CML & BPRMF & NGCF & EMCDR & SSCDR & TMCDR & SA-VAE & CDRIB & UniCDR & \method{} \\ 
 \hline
\multirow{2}{*}{Sport} & HR & 5.82$\pm$0.20 & 5.75$\pm$0.26 & 7.22$\pm$0.11 & 7.41$\pm$0.16 & 7.27$\pm$0.02 & 7.18$\pm$0.07 & 7.51$\pm$0.02 & \textbf{11.10$\pm$0.29} & 10.67$\pm$0.19 & {\underline{10.72$\pm$0.32}} \\
 & NDCG & 3.29$\pm$0.16 & 3.16$\pm$0.15 & 3.63$\pm$0.07 & 4.03$\pm$0.12 & 3.75$\pm$0.02 & 3.84$\pm$0.04 & 3.72$\pm$0.02 & 5.78$\pm$0.15 & \textbf{6.19$\pm$0.17} & {\underline{5.81$\pm$0.26}} \\ \cline{2-12} 
\multirow{2}{*}{Cloth} & HR & 6.97$\pm$0.11 & 6.75$\pm$0.13 & 7.07$\pm$0.30 & 7.91$\pm$0.15 & 6.12$\pm$0.05 & 8.11$\pm$0.16 & 7.21$\pm$0.05 & 12.23$\pm$0.24 & {\underline{12.28$\pm$0.24}} & \textbf{12.85$\pm$0.42} \\
 & NDCG & 3.92$\pm$0.14 & 3.26$\pm$0.15 & 3.48$\pm$0.13 & 5.17$\pm$0.08 & 3.06$\pm$0.04 & 5.05$\pm$0.12 & 4.59$\pm$0.08 & 6.79$\pm$0.22 & \textbf{7.31$\pm$0.26} & {\underline{6.92$\pm$0.32}} \\ 
 \hline\hline
\multirow{2}{*}{Game} & HR & 2.82$\pm$0.18 & 3.77$\pm$0.40 & 5.14$\pm$0.22 & 5.07$\pm$0.17 & 3.48$\pm$0.06 & 5.36$\pm$0.09 & 5.84$\pm$0.13 & {\underline{8.72$\pm$0.41}} & 7.68$\pm$0.32 & \textbf{9.54$\pm$0.28} \\
 & NDCG & 1.44$\pm$0.09 & 1.89$\pm$0.19 & 2.73$\pm$0.09 & 2.44$\pm$0.08 & 1.59$\pm$0.03 & 2.58$\pm$0.07 & 2.78$\pm$0.06 & 4.58$\pm$0.20 & {\underline{4.63$\pm$0.17}} & \textbf{4.87$\pm$0.27} \\ \cline{2-12} 
\multirow{2}{*}{Video} & HR & 3.07$\pm$0.10 & 4.46$\pm$0.56 & 7.41$\pm$0.18 & 8.43$\pm$0.04 & 5.51$\pm$0.08 & 8.85$\pm$0.11 & 7.46$\pm$0.13 & {\underline{12.66$\pm$0.39}} & 10.32$\pm$0.36 & \textbf{13.38$\pm$0.31} \\
 & NDCG & 1.30$\pm$0.08 & 2.36$\pm$0.34 & 3.87$\pm$0.10 & 4.29$\pm$0.02 & 2.61$\pm$0.02 & 4.41$\pm$0.08 & 3.71$\pm$0.06 & {\underline{6.66$\pm$0.20}} & 5.43$\pm$0.16 & \textbf{6.86$\pm$0.14} \\ 
 \hline\hline
\multirow{2}{*}{Music} & HR & 8.70$\pm$0.27 & 8.74$\pm$0.13 & 8.86$\pm$0.10 & 8.95$\pm$0.12 & 3.59$\pm$0.04 & 9.48$\pm$0.08 & 8.57$\pm$0.15 & {\underline{14.72$\pm$0.23}} & 11.59$\pm$0.35 & \textbf{15.92$\pm$0.21} \\
 & NDCG & 4.53$\pm$0.14 & 4.72$\pm$0.04 & 4.15$\pm$0.09 & 4.70$\pm$0.08 & 1.82$\pm$0.01 & 5.15$\pm$0.05 & 4.48$\pm$0.06 & {\underline{7.98$\pm$0.10}} & 6.08$\pm$0.22 & \textbf{8.56$\pm$0.13} \\ \cline{2-12} 
\multirow{2}{*}{Movie} & HR & 7.87$\pm$0.11 & 9.12$\pm$0.15 & 9.92$\pm$0.16 & 11.89$\pm$0.10 & 5.47$\pm$0.03 & 10.22$\pm$0.15 & 11.51$\pm$0.18 & {\underline{14.86$\pm$0.44}} & 12.40$\pm$0.45 & \textbf{16.33$\pm$0.28} \\
 & NDCG & 3.95$\pm$0.03 & 4.72$\pm$0.08 & 4.53$\pm$0.10 & 6.06$\pm$0.14 & 2.72$\pm$0.01 & 6.00$\pm$0.04 & 5.66$\pm$0.11 & {\underline{7.63$\pm$0.23}} & 6.41$\pm$0.22 & \textbf{8.39$\pm$0.17} \\ 
 \hline\hline
\multirow{2}{*}{Phone} & HR & 11.56$\pm$0.23 & 11.98$\pm$0.22 & 13.62$\pm$0.20 & 13.84$\pm$0.07 & 6.14$\pm$0.03 & 13.95$\pm$0.22 & 14.35$\pm$0.29 & {\underline{18.37$\pm$0.32}} & 14.27$\pm$0.37 & \textbf{18.74$\pm$0.31} \\
 & NDCG & 6.41$\pm$0.13 & 7.65$\pm$0.07 & 7.47$\pm$0.12 & 7.73$\pm$0.03 & 3.33$\pm$0.01 & 7.56$\pm$0.09 & 7.66$\pm$0.18 & {\underline{10.16$\pm$0.25}} & 8.87$\pm$0.25 & \textbf{10.19$\pm$0.22} \\ \cline{2-12} 
\multirow{2}{*}{Elec} & HR & 12.44$\pm$0.10 & 12.78$\pm$0.33 & 16.24$\pm$0.21 & 17.01$\pm$0.03 & 10.34$\pm$0.03 & 16.38$\pm$0.08 & 17.21$\pm$0.19 & {\underline{20.96$\pm$0.31}} & 15.67$\pm$0.47 & \textbf{21.03$\pm$0.29} \\
 & NDCG & 6.85$\pm$0.12 & 7.46$\pm$0.23 & 9.09$\pm$0.11 & 9.78$\pm$0.02 & 5.50$\pm$0.01 & 9.24$\pm$0.05 & 9.49$\pm$0.13 & \textbf{12.01$\pm$0.22} & 9.25$\pm$0.30 & {\underline{11.54$\pm$0.23}}\\
\hline\hline
\end{tabular}%
}
\caption{Performance comparison (expressed in \%) of CDR on cross domain recommendations. The best performance is \textbf{bold-faced} and the runner-up is \underline{underlined} in terms of the corresponding metric.}
\label{tab:experiments}
\vspace{-0.4cm}
\end{table*}


\subsection{Experimental Setup}
\subsubsection{Datasets.}
We experiment on four domain pairs from the public Amazon dataset~\footnote{{\url{http://jmcauley.ucsd.edu/data/amazon/index_2014.html}}}, namely music-movie, phone-electronic, 
cloth-sport, and game-video. We filter out the users and items with fewer than $5$ and $10$ interactions in their respective domains to improve data quality.
Following prior works~\cite{cao2022cross, cao2023towards}, we randomly select $20\%$ of overlapping users (i.e., those observed in both source and target domains) and treat them as cold-start users by removing their target domain interactions during testing and validation, using the remaining users for training. 

\subsubsection{Baselines.}
We compare \method{} with  two baseline groups: 
(A) Single-domain recommendation, where interactions from both domains are merged into one and standard CF-based methods are applied, including CML~\cite{hsieh2017collaborative}, BPR-MF~\cite{rendle2012bpr}, and NGCF~\cite{wang2019neural}. 
(B) Cross-domain recommendation (CDR) models follow two paradigms. 
The first, Embedding and Mapping approach for CDR (EMCDR)~\cite{man2017cross} pre-trains single-domain models (e.g., CML, BPR-MF, NGCF) to initialize user/item representations and includes extensions like SSCDR~\cite{kang2019semi}, TMCDR~\cite{zhu2021transfer}, and SA-VAE~\cite{salah2021towards}. 
The second, disentangled representation learning encodes domain-shared and domain-specific knowledge, as seen in CDRIB~\cite{cao2022cross} and UniCDR~\cite{cao2023towards}.


\subsubsection{Evaluation.}
We employ the leave-one-out approach to evaluate the performance of all methods.
Given a ground truth interaction $(u, v)$ in target domain $\mathcal{T}$, we randomly select $999$ items 
from the item set $\mathcal{V}^{\mathcal{T}}$ as negative samples. We then generate $1{,}000$ records ($1$ positive and $999$ negative samples) using the learned representation $z^s_u$ from source domain $\mathcal{S}$ and randomly-selected positive sample $z^t_v$ or negative sample $z^t_{v'}$ from the target domain $\mathcal{T}$.
We rank the record list and use two widely-used recommendation evaluation metrics, namely NDCG@10 and HR@10, to evaluate the top-$10$ recommendation performance.

\smallskip\noindent\textbf{Implement Details.}
In our experiments, we set the embedding dimension to $128$, the batch size to $1{,}024$, and the slope of LeakyReLU to $0.05$. In specific, we tune the number of the graph encoder layer $L$ in the range of $[1, 6]$, the number of latent factors $K$ in the range of $[1, 6]$, and the values of ${\beta}$ and ${\lambda}$ of the objective function in the range of $[0, 0.5]$. Additionally, we turn the dropout rate in the range $[0, 0.5]$, and the learning rate in the range of $[0, 0.005]$.
For all methods, we tune the hyper-parameters by grid search, run each experiment five times with different random seeds and record the best result of each time according to the HR@10 performance on the validation set.


\subsection{Performance Analysis}

Table~\ref{tab:experiments} presents the mean best recommendation performance for all methods evaluated across four CDR scenarios, and we have several key observations:
(1) Graph-based approaches like NGCF consistently outperform methods CML and BPR-MF, highlighting the effectiveness of graph convolution in capturing multi-hop neighborhood and high-order interaction information.
(2) Compared to the single-domain approaches, cross-domain methods show a superior performance, which demonstrates the ability of the embedding and mapping paradigm to capture domain-specific knowledge and nuances between source and target domains.
(3) Compared to cross-domain approaches,  \method{} excels on the game-video and music-movie datasets and remains competitive on the sport-cloth and phone-electronic datasets when compared to state-of-the-art graph-encoder models such as CDRIB and UniCDR.
Additionally, \method{} outperforms all other EMCDR-based methods, indicating the effectiveness of intra-domain and inter-domain contrast mechanisms to filter out irrelevant collaborative information.

\begin{table}[t]
\setlength{\tabcolsep}{1pt}
%
\resizebox{\linewidth}{!}{
\begin{tabular}{ccccccc}
\hline\hline
\multirow{2}{*}{D} & \multirow{2}{*}{M} & \multicolumn{5}{c}{Model Variants} \\ \cline{3-7} 
 &  & Variant1 & Variant2 & Variant3 & Variant4 & \method{} \\ \hline
\multirow{2}{*}{Sport} & H & 9.81$\pm$0.24 & 10.64$\pm$0.19 & 10.62$\pm$0.37 & 9.56$\pm$0.27 & \textbf{10.72$\pm$0.32} \\
 & N & 4.81$\pm$0.13 & 5.40$\pm$0.15 & 5.38$\pm$0.29 & 4.31$\pm$0.25 & \textbf{5.81$\pm$0.26} \\ \cline{2-7} 
\multirow{2}{*}{Cloth} & H & 11.79$\pm$0.29 & 12.55$\pm$0.20 & 12.2$\pm$0.26 & 11.88$\pm$0.31 & \textbf{12.85$\pm$0.42} \\
 & N & 6.14$\pm$0.21 & 6.69$\pm$0.22 & 6.16$\pm$0.17 & 5.27$\pm$0.26 & \textbf{6.92$\pm$0.32} \\ 
 \hline\hline
\multirow{2}{*}{Game} & H & 9.20$\pm$0.32 & 9.29$\pm$0.35 & 9.41$\pm$0.24 & 8.62$\pm$0.26 & \textbf{9.54$\pm$0.28} \\
 & N & 4.33$\pm$0.25 & 4.53$\pm$0.29 & 4.73$\pm$0.22 & 3.97$\pm$0.19 & \textbf{4.87$\pm$0.27} \\ \cline{2-7} 
\multirow{2}{*}{Video} & H & 11.5$\pm$0.31 & 12.06$\pm$0.33 & 12.26$\pm$0.29 & 12.4$\pm$0.23 & \textbf{13.38$\pm$0.31} \\
 & N & 6.14$\pm$0.15 & 6.29$\pm$0.19 & 6.38$\pm$0.20 & 6.35$\pm$0.12 & \textbf{6.86$\pm$0.14} \\ 
 \hline\hline
\multirow{2}{*}{Music} & H & 14.96$\pm$0.23 & 15.14$\pm$0.28 & 13.51$\pm$0.30 & 13.87$\pm$0.25 & \textbf{15.92$\pm$0.21} \\
 & N & 7.65$\pm$0.08 & 7.75$\pm$0.20 & 7.12$\pm$0.17 & 7.19$\pm$0.14 & \textbf{8.56$\pm$0.13} \\ \cline{2-7} 
\multirow{2}{*}{Movie} & H & 14.81$\pm$0.22 & 15.42$\pm$0.33 & 14.33$\pm$0.26 & 14.12$\pm$0.33 & \textbf{16.33$\pm$0.28} \\
 & N & 7.49$\pm$0.17 & 7.87$\pm$0.19 & 7.31$\pm$0.20 & 7.33$\pm$0.24 & \textbf{8.39$\pm$0.17} \\ 
 \hline\hline
\multirow{2}{*}{Phone} & H & 17.27$\pm$0.29 & 18.21$\pm$0.35 & 18.22$\pm$0.30 & 17.70$\pm$0.34 & \textbf{18.74$\pm$0.31} \\
 & N & 8.82$\pm$0.23 & 9.44$\pm$0.26 & 10.01$\pm$0.18 & 9.58$\pm$0.26 & \textbf{10.19$\pm$0.22} \\ \cline{2-7} 
\multirow{2}{*}{Elec} & H & 18.06$\pm$0.32 & 18.65$\pm$0.30 & 19.56$\pm$0.24 & 19.48$\pm$0.30 & \textbf{21.03$\pm$0.29} \\
 & N & 9.73$\pm$0.27 & 10.73$\pm$0.25 & 10.89$\pm$0.19 & 10.96$\pm$0.25 & \textbf{11.54$\pm$0.23} \\ 
\hline\hline
\end{tabular}
}%
\caption{Ablation studies on the variants of \method{}. Dataset, Metric@10, HR, NDCG are abbreviated as D, M, H, N.}
\label{tab:ablation}
\vspace{-0.4cm}
\end{table}

\begin{table*}[htbp]
\resizebox{\textwidth}{!}{
\begin{tabular}{c|ccccc}
\hline\hline
Setting & \multicolumn{5}{c}{Five Favorite Preference} \\ \hline
\multirow{2}{*}{\begin{tabular}[c]{@{}c@{}}Source Domain\\ Training\end{tabular}} & $u_1$: Home Entertainment & Studio Specials & Art House \& International & Action & Science Fiction \\
 & $u_2$: Home Entertainment & Studio Specials & Art House \& International & Drama & Fantasy \\ \hline
\multirow{2}{*}{\begin{tabular}[c]{@{}c@{}}Target Domain \\ Test Groud Truth\end{tabular}} & $u_1$: Pop & Metal & World Music & Country & Dance \& Electronic \\
 & $u_2$: Pop & Oldies & Comedy \& Spoken Word & Christian & Soundtracks \\ 
 \hline\hline
\multirow{2}{*}{\begin{tabular}[c]{@{}c@{}}Target Domain\\ Prediction by CDRIB\end{tabular}} & $u_1$: \textbf{Pop} & Rock & Folk & \textbf{Country} & Movie Soundtracks \\
 & $u_2$: \textbf{Pop} & Rock & Roadway & Country & Instructional \\ \hline
\multirow{2}{*}{\begin{tabular}[c]{@{}c@{}}Target Domain\\ Prediction by DisCo\end{tabular}} & $u_1$: \textbf{Pop} & Rock & New Age & \textbf{Country} & \textbf{Dance \& Electronic} \\
 & $u_2$: \textbf{Pop} & Jazz & \textbf{Comedy \& Spoken Word} & Classical & Symphonies \\ 
 \hline\hline
\end{tabular}
}
\caption{Case Study on a user pair ($u_1,u_2$). Predicting results are \textbf{bold-faced} if they match the target domain test ground truth.}
\label{tab:case}
\vspace{-5pt}
\end{table*}

\subsection{Ablation Study}
To evaluate the effectiveness of the various modules in \method{} and understand their functions, we compare \method{} with four variants:
(1) Variant 1: It eliminates the cross-domain decoder.
(2) Variant 2: It sets a uniform distribution of latent intents, with $p_(k|u_i) = 1 / K$.
(3) Variant 3: It omits the intra-domain contrast loss $\mathcal{L}_{orth}$, allowing the $k$ intents to be non-orthogonal.
(4) Variant 4: It removes affinity graph construction and multi-step random walks ( $\tilde{R}^d = I$).

The results of \method{} and its variants are shown in Table~\ref{tab:ablation}. We find that: (1) Compared to Variant 1, a notable decline in performance underscores the specialized knowledge across domains. This is attributed to the removal of the mapping function, which impairs the ability to learn domain-specific knowledge.
(2) The comparison with Variant 2 reveals that individual users have distinct preference weights across domains, and assigning equal weight to each intent can obscure vital user preference information.
(3) The comparison with Variant 3 demonstrates that maintaining orthogonality among the $K$ intents improves prediction accuracy by capturing more diverse and richer user preference information.
(4) Compared to Variant 4, the affinity graph construction and multi-step random walks leverage the inherent similarity structure of users, enhancing both intra-domain and inter-domain contrasts to filter out irrelevant collaborative information and avoid the negative transfer problem.
\begin{figure}[t]
\centering
\begin{subfigure}{0.49\linewidth}
    \includegraphics[width=\linewidth]{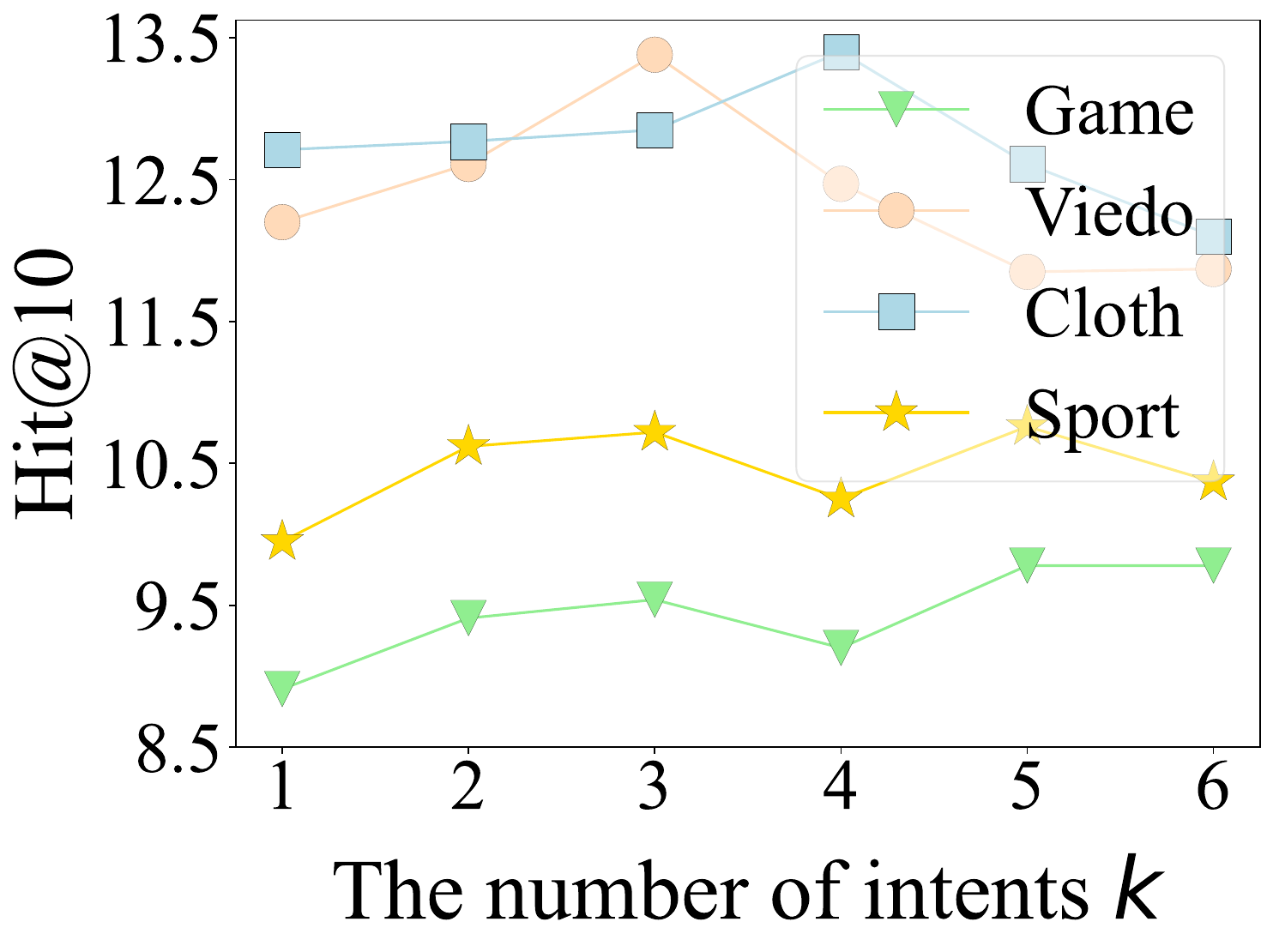}
    \caption{Hit@10}
    \label{fig:3_a}
\end{subfigure}
\begin{subfigure}{0.49\linewidth}
    \includegraphics[width=\linewidth]{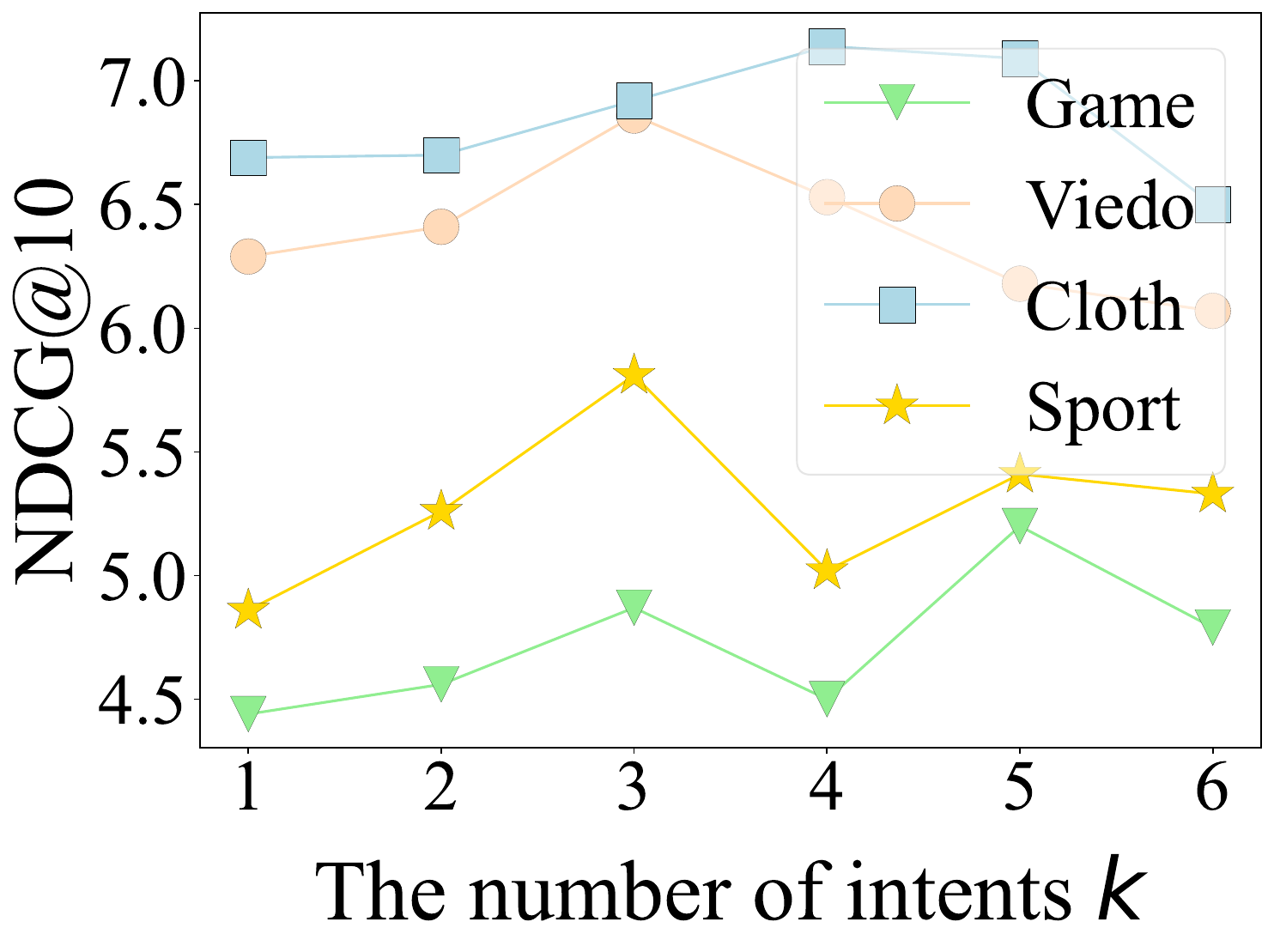}
    \caption{NDCG@10}
    \label{fig:3_b}
\end{subfigure}
\caption{Performance comparison w.r.t. different numbers of user intent $K$.}
\label{fig:intents}
\end{figure}

\begin{figure}[t]
\vspace{-0.4cm}
\centering
\begin{subfigure}{0.49\linewidth}
    \includegraphics[width=\linewidth]{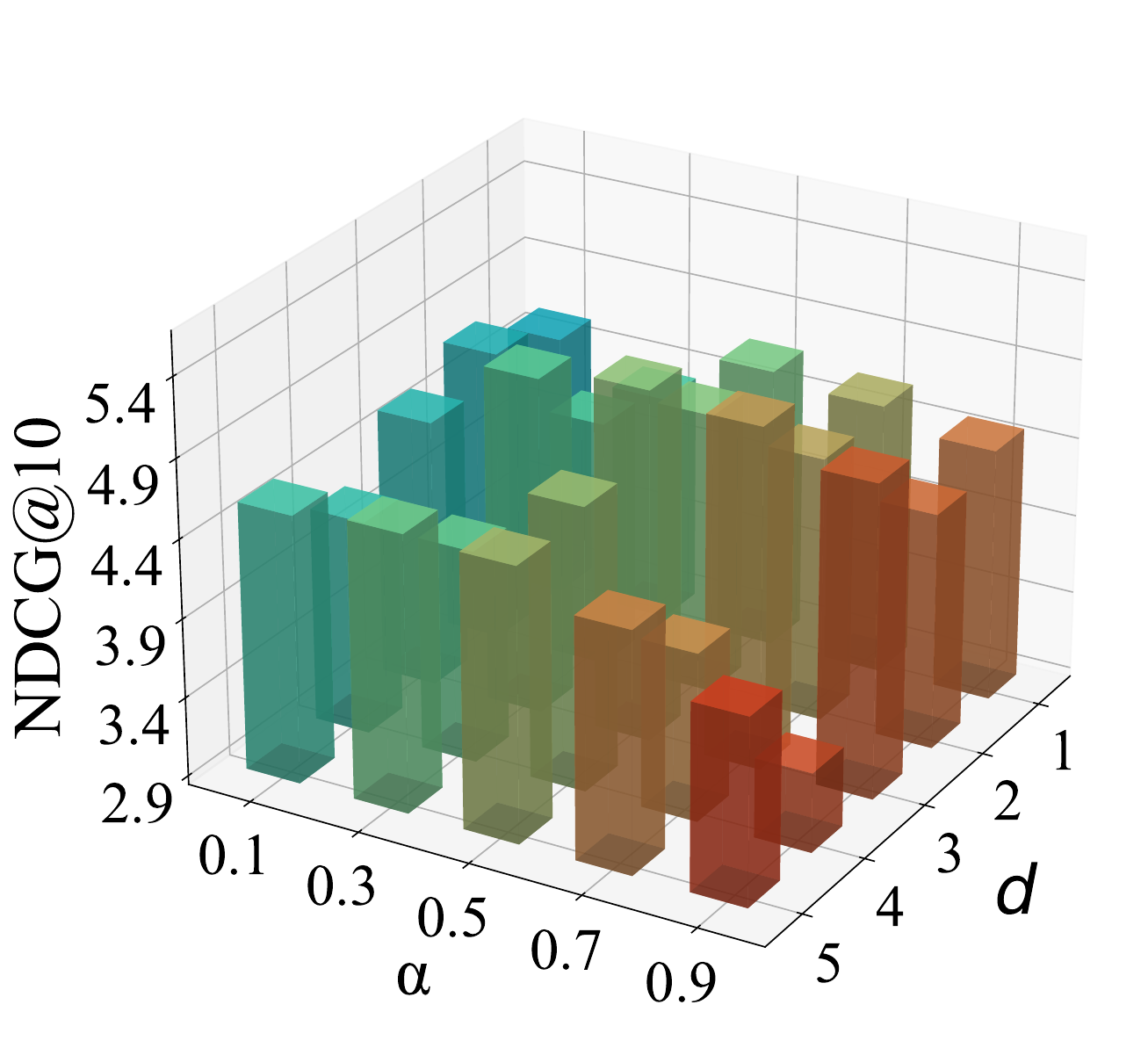}
    \caption{Game}
    \label{fig:4_a}
\end{subfigure}
\begin{subfigure}{0.49\linewidth}
    \includegraphics[width=\linewidth]{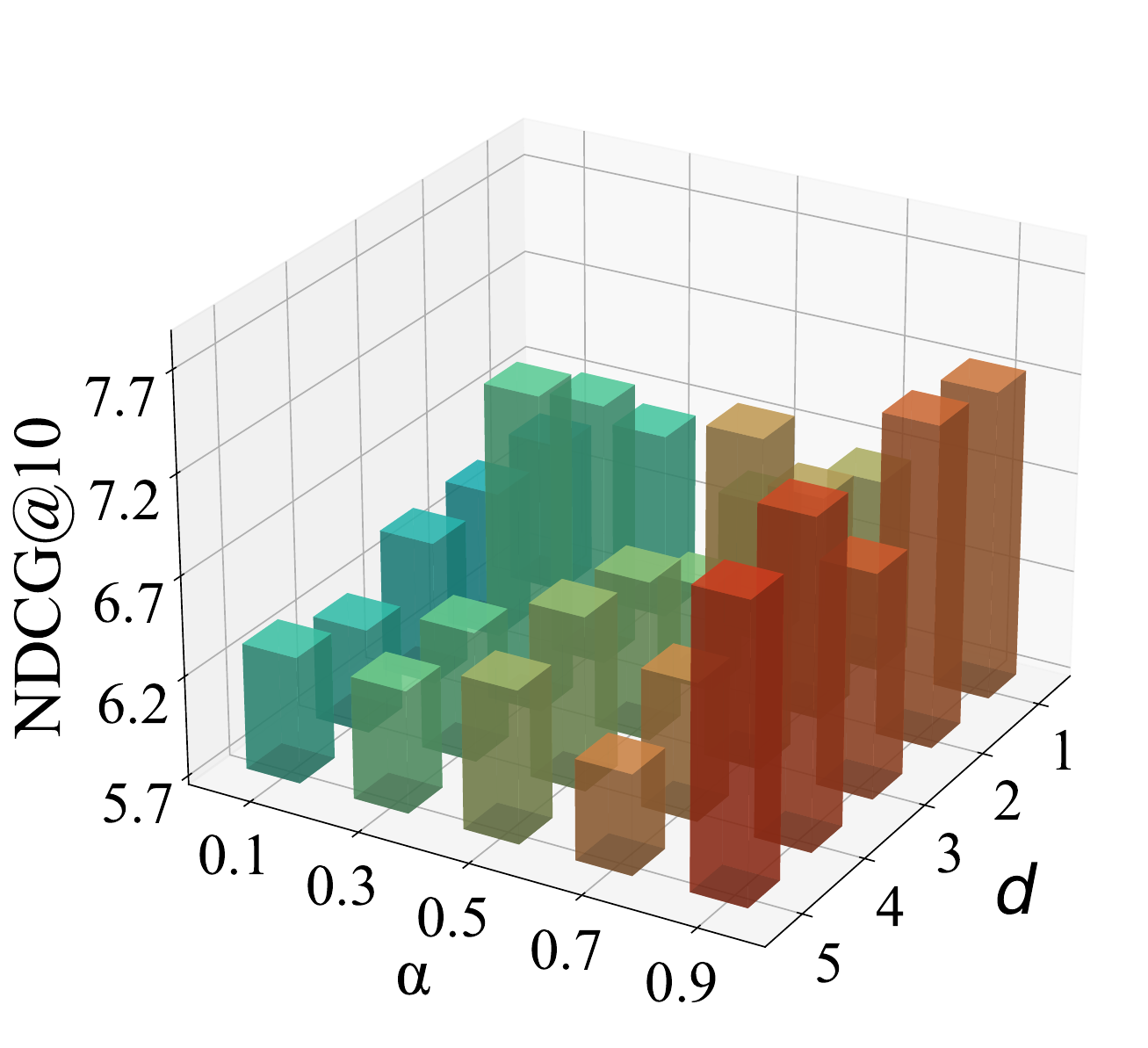}
    \caption{Video}
    \label{fig:4_b}
\end{subfigure}
\caption{Performance comparison w.r.t. different values of $d$ and $\alpha$ for random walks and high-order user similarity.}
\vspace{-0.4cm}
\label{fig:similarity}
\end{figure}
\begin{figure}[t]
\centering
\begin{subfigure}{0.49\linewidth}
    \includegraphics[width=\linewidth]{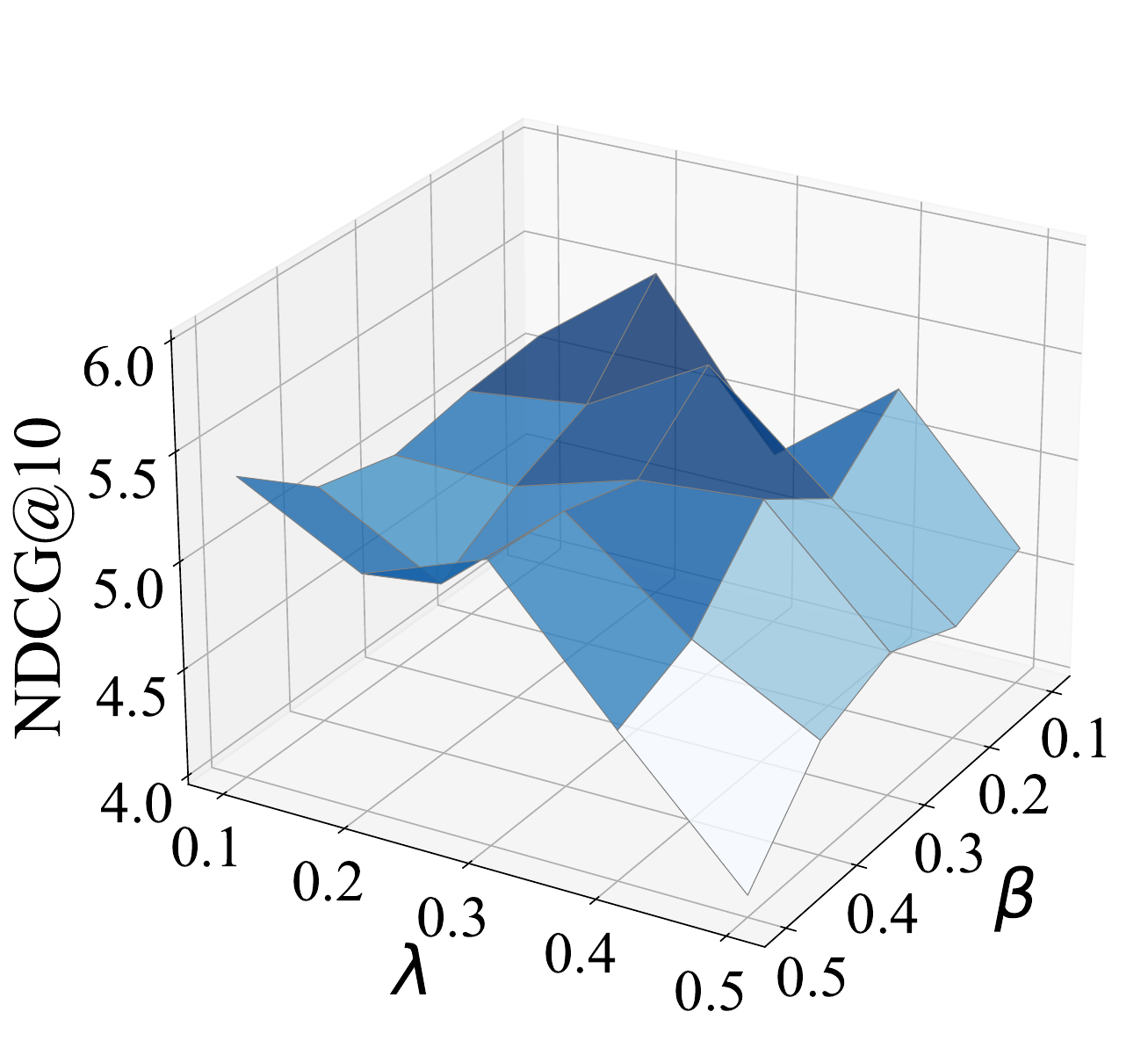}
    \caption{Sport}
    \label{fig:5_a}
\end{subfigure}
\begin{subfigure}{0.49\linewidth}
    \includegraphics[width=\linewidth]{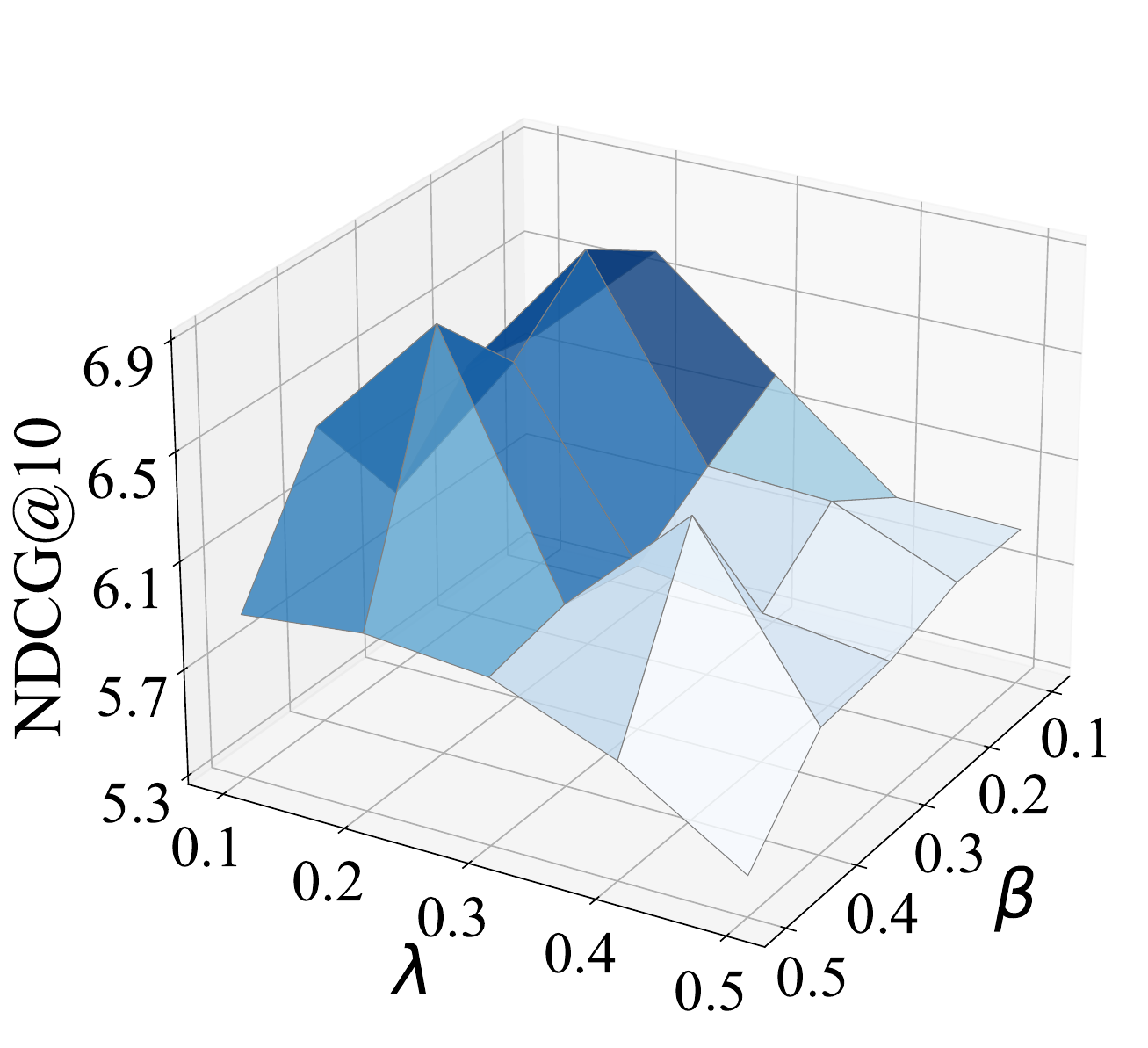}
    \caption{Cloth}
    \label{fig:5_b}
\end{subfigure}
\caption{Performance comparison w.r.t. different values of $\lambda$ and $\beta$ for the overall objective.}
\vspace{-0.5cm}
\label{fig:objective}
\end{figure}

\subsection{Parameter Analysis}
We analyze the impact of three groups of hyper-parameters in \method{}: the number of latent intents, the parameter pair $\alpha$ and $d$ in generating user similarity $T_k$ (Equation~\ref{randomWalk}), and the multiplier factors $\beta$ and $\lambda$ in the objective function (Equation~\ref{objectiveFunc}). We find that:
(1) The optimal number of latent intents is closely related to the specific domain. As shown in Figure~\ref{fig:intents}, scenarios with higher interaction generally exhibit a larger number of user intents. For instance, in the domains of video and sport with dense interactions, a larger number of user intents improves performance when transferring knowledge to cold-start users in game and cloth scenarios.
(2) Figure~\ref{fig:similarity} demonstrates the influence of parameters $\alpha$ and $d$ on generating the user similarity matrix $T_k$ through multi-step random walks. 
Our results show that higher-order random walks (user similarity) generally outperform lower-order ones.
Additionally, The \method{} performs well when $d$ is set between $3$ to $5$ steps.
(3) Figure~\ref{fig:objective} shows that an excessively high contrast loss factor $\lambda$ reduces performance. And the optimal $\beta$ is usually between $0.2$ and $0.4$.

\subsection{Case Study}
We conduct a case study to comfirm that \method{} effectively mitigates negative transfer. 
As shown in Table~\ref{tab:case}, two overlapping users ($u_1$,$u_2$) from the music-movie dataset are selected, with movies as the source domain and music as the target domain. The five most popular tags from movie interactions represent users' source domain preferences. 
Similarly, the five most popular tags from the top $10$ ranked items generated by model depict their target domain preferences.
In the target domain test ground truth, $u_1$ and $u_2$ share similar preferences in the source domain but differ in the target domain, illustrating the negative transfer issue.
The results show that \method{} mitigates negative transfer, in contrast to the cold start CDR baseline, CDRIB, which fails to filter out irrelevant source-domain collaborative information.
	
 


\section{Conclusion}
In this paper, we propose a novel framework termed \method{} for cold-start cross-domain recommendation tasks. 
To address the negative transfer issue, we argue that users with similar preferences in the source domain may have different interests in the target domain, emphasizing the need to capture diverse user intents to filter irrelevant collaborative information. 
Specifically, a multi-channel graph encoder is employed to extract diverse user intents.
Then, multi-step random walks on affinity graphs learn high-order user similarities under each intent in the source and target domains. 
Additionally, we treat one of the domains as the target domain and propose an intent-wise disentangled contrastive learning approach to retain user similarity relationships while explicitly identifying the rationale between the source and target domains. 
Experimental results on four benchmark datasets demonstrate the efficacy of our \method{}. 

\section{Acknowledgments}
This paper is partially supported by the National Natural Science Foundation of China with Grant Numbers 62276002 and 62306014 as well as the Fundamental Research Funds for the Central Universities in UIBE (Grant No. 23QN02).



\bibliography{aaai25}
\end{document}


\maketitle


%

\appendix



\section{Training Details}
\subsection{Hardware and Software Configurations}
We conduct the experiments with:
\begin{itemize}[leftmargin=*]
    \item Operating System: Ubuntu 18.04.6 LTS
    \item CPU: Intel(R) Xeon(R) CPU E5-2697 v4 @ 2.30GHz
    \item GPU: NVIDIA GeForce RTX 3090
    \item Software: Python 3.8.19; NumPy 1.24.4; PyTorch 1.10.1+cu111; PyTorch Geometric 2.5.3
\end{itemize}

\subsection{Datasets}
We conduct experiments on public Amazon dataset \cite{ni2019justifying}, which is available at \url{http://jmcauley.ucsd.edu/data/amazon/index_2014.html} with an unspecified license. 
This dataset includes product reviews and metadata from Amazon, encompassing 24 disjoint item domains. 
We select 8 domains and construct four pairs of CDR scenarios for our experiments: Music-Movie, Phone-Electronic, Cloth-Sport, and Game-Video.
Following previous works ~\cite{cao2022cross, cao2023towards}, items with fewer than 10 interactions and users with fewer than 5 interactions in their respective domains are filtered out, ensuring representative embeddings could be learned from their domain interactions. 
The goal of CDR is to recommend items in the target domain for users (cold-start users) who are only observed in the source domain. 
Consequently, nealy about 20\% of overlapping users are selected as cold-start users (e.g., 10\% from Sport to recommend in Cloth and the remaining 10\% from Cloth to recommend in Sport) for both the test and validation sets, while the remaining users are used for training. 
The statistics for the four CDR scenarios are provided in Table ~\ref{tab:statistics}.

\begin{table*}[]
\caption{Statistics of four CDR scenarios. (\#Overlap denotes the number of overlapping users in the training set. \#Density denotes the proportion of interactions within the user-item matrix.)}
\centering
\label{tab:statistics}
\begin{tabular}{c|cc|cc|ccc|c}
\hline\hline
Scenarios  & $|\mathcal{U}|$    & $|\mathcal{V}|$   &Training   &\#Overlap      &Validation     &Test       &Cold-start     & \#Density                           \\ \hline
Sport      & $27{,}328$  & $12{,}655$  & $163{,}291$  &\multirow{2}{*}{$7{,}857$}  &$3{,}589$  &$3{,}546$  &$981$   & $0{.}47$\textperthousand \\
Cloth      & $41{,}829$  & $17{,}943$  & $187{,}880$  &  &$3{,}156$  &$3{,}085$  &$990$   &$0{.}25$\textperthousand \\ \hline\hline
Game       & $25{,}025$  & $12{,}3139$ & $155{,}036$  &\multirow{2}{*}{$1{,}737$}  &$1{,}381$  &$1{,}304$  &$226$   & $0{.}51$\textperthousand \\
Video      & $19{,}457$  & $8{,}751$   & $156{,}091$  &  &$1{,}435$  &$1{,}458$  &$217$&  $0{.}93$\textperthousand \\ \hline\hline
Music      & $50{,}841$  & $43{,}858$  & $674{,}233$     &\multirow{2}{*}{$15{,}081$}  &$19{,}837$  &$19{,}670$  &$1{,}893$  & $0{.}30$\textperthousand \\
Movie      & $87{,}875$  & $38{,}643$  & $1{,}127{,}424$ &  &$28{,}589$  &$28{,}876$  &$1{,}885$  & $0{.}33$\textperthousand \\ \hline\hline
Phone      & $27{,}519$  & $9{,}481$   & $148{,}271$  &\multirow{2}{*}{$16{,}337$}  &$6{,}417$  &$6{,}322$  &$2{,}049$   & $0{.}57$\textperthousand \\
Electronic & $107{,}984$ & $40{,}460$  & $821{,}301$  &  &$15{,}199$ &$15{,}053$  &$2{,}042$   & $0{.}19$\textperthousand \\ \hline\hline
\end{tabular}

\end{table*}

\subsection{Baselines}
Two groups of baselines are compared with DisCo: single-domain recommendation models and cross-domain recommendation models.

For single-domain recommendation models, we select Collaborative Metric Learning (CML) ~\cite{hsieh2017collaborative}, Bayesian Personalized Ranking Matrix Factorization (BPR-MF) ~\cite{rendle2012bpr}, and Neural Graph Collaborative Filtering (NGCF) ~\cite{wang2019neural} as baselines. 
These methods are primarily CF-based. 
CML integrates metric learning with collaborative filtering to capture nuanced user preferences and item relationships.
BPR-MF optimizes personalized ranking in recommender systems using implicit feedback data and is a variant of Matrix Factorization (MF). 
NGCF uses a graph neural network approach to capture collaborative signals by propagating embeddings along the user-item interaction graph. 
When applying single-domain methods to cross-domain scenarios, we merge all interactions from both domains into a single domain.

For cross-domain recommendation models, there are primarily two paradigms.
The first paradigm involves the Embedding and Mapping approach for CDR (EMCDR) ~\cite{man2017cross}, which involves learning separate user and item representations for each domain independently and then projecting user embeddings from the source domain to the target domain. 
For EMCDR-base method, we select Semi-Supervised Learning for CDR (SSCDR) ~\cite{kang2019semi}, Transfer-meta framework for CDR (TMCDR) ~\cite{zhu2021transfer} as baselines.
SSCDR enhances the mapping process with semi-supervised learning techniques.
TMCDR incorporates meta-learning for better adaptation across domains.
The second paradigm focuses on learning an unbiased representation that directly encodes domain-invariant knowledge without relying on explicit mapping functions between domains. 
For this paradigm, we select Source-Aligned Variational Models for CDR (SA-VAE) ~\cite{salah2021towards}, CDR via Variational Information Bottleneck (CDRIB) \cite{cao2022cross} and Universal CDR (UniCDR) \cite{cao2023towards} as baselines.
SA-VAE uses variational autoencoders to align the hidden space between domains to better capture shared information.
CDRIB uses information bottleneck principles to filter out domain-specific noise and focus on shared information.
UniCDR builds a unified representation space that can effectively generalize across inter-domain and intra-domain CDR tasks.

\subsection{Implementation details}
We implement our model in PyTorch. We use the Adam optimizer, a variant of Stochastic Gradient Descent (SGD) with adaptive moment estimation. GCN is employed as the message-passing layers. Since the number of latent intents is unknown, we search for the optimal number of intents $K$ from 1 to 6. 
To generate contrastive representations, we use two Siamese encoders: an online encoder $f_q$ and a momentum-based target encoder $f_k$, to produce dual embeddings of users and items. 
The target encoder shares the same architecture with the online encoder and is updated using the Exponential Moving Average (EMA) of $f_q$ instead of backpropagation. 
We do not employ existing graph augmentation methods, as they may fail to preserve the original semantic information of the nodes.
The learning rate is fixed at 0.004, as a lower learning rate may lead to local optima, while a higher learning rate can hinder convergence. 
The number of GCN layers is also critical, and 4 layers of GCNs strike a balance, providing good representations of users and items without causing over-smoothing. 
A dropout rate between 0.3 and 0.4 typically yields good performance. 
Other hyperparameters are optimized through grid search.

We list the detailed training procedure of our method in Algorithm \ref{algorithm:pesudo}.

\begin{algorithm}[h]
\small
\caption{The overall training process of \method{}}
\label{algorithm:pesudo}
\flushleft{\textbf{Input}: Observed data $\mathcal{G}_{\mathcal{S}}$ and $\mathcal{G}_\mathcal{T}$ of source and target domains.} \\
\flushleft{$\textbf{Output}$: Cold-start prediction model $\mathcal{F}$ with parameter $\Theta$.} \\
\begin{algorithmic}[1]
\WHILE{not convergence}
\FOR{$\mathcal{G}\in\{\mathcal{G}_{\mathcal{S}},\mathcal{G}_{\mathcal{T}}\}$}
\STATE Encode $\mathcal{G}$ with two Siamese user intent graph encoder (online and target encoder) \tcp*[r]{Eq. 1-2}
\STATE Construct $R$ with target encoder output and perform multi-step random walks to obtain $T_k$. \tcp*[r]{Eq. 3-4}
\tcc{Intra-domain Contrast}
\STATE Get intra-domain contrastive loss $\mathcal{L}_{intra}$ \tcp*[r]{Eq. 5}
\STATE Obtain orthogonality loss $\mathcal{L}_{orth}$ \tcp*[r]{Eq. 7}
\ENDFOR \\
\tcc{Inter-domain Contrast}
\STATE Cross-domain information decoding \tcp*[r]{Eq. 8}
\STATE Get inter-domain contrastive loss $\mathcal{L}_{inter}$ \tcp*[r]{Eq. 9}
\STATE Adapt user intent and get prediction loss $\mathcal{L}_{rec}^\mathcal{S}$, $\mathcal{L}_{rec}^\mathcal{T}$ \tcp*[r]{Eq. 16-18}
\STATE Compute the overall objective and update; \tcp*[r]{Eq. 19}
\ENDWHILE
\STATE \textbf{return} $\mathcal{F}$ with parameter $\Theta$
\end{algorithmic}
\end{algorithm}

\section{Ablation studies on Siamese encoders}
\begin{table}[]
\centering
\caption{Ablation studies on siamese encoder variants of DisCo.}
\label{tab:siamese}
\resizebox{\columnwidth}{!}{%
\begin{tabular}{ccccc}
\hline\hline
\multirow{2}{*}{Dataset} & \multirow{2}{*}{Metric@10} & \multicolumn{3}{c}{Model Variants} \\ \cline{3-5} 
 &  & Variant1 & Variant2 & DisCo \\ \hline\hline
\multirow{2}{*}{Sport} & Hit@10 & 9.95$\pm$0.35 & 10.23$\pm$0.29 & \textbf{10.72$\pm$0.32} \\
 & NDCG@10 & 5.11$\pm$0.27 & 5.31$\pm$0.21 & \textbf{5.81$\pm$0.26} \\ \cline{2-5} 
\multirow{2}{*}{Cloth} & Hit@10 & 10.27$\pm$0.29 & 11.69$\pm$0.33 & \textbf{12.85$\pm$0.42} \\
 & NDCG@10 & 5.52$\pm$0.17 & 6.36$\pm$0.27 & \textbf{6.92$\pm$0.32} \\ \hline\hline
\multirow{2}{*}{Game} & Hit@10 & 7.52$\pm$0.16 & 9.34$\pm$0.25 & \textbf{9.54$\pm$0.28} \\
 & NDCG@10 & 3.66$\pm$0.11 & 4.48$\pm$0.22 & \textbf{4.87$\pm$0.27} \\ \cline{2-5} 
\multirow{2}{*}{Video} & Hit@10 & 11.43$\pm$0.18 & 12.89$\pm$0.26 & \textbf{13.38$\pm$0.31} \\
 & NDCG@10 & 5.92$\pm$0.12 & 6.60$\pm$0.13 & \textbf{6.86$\pm$0.14} \\ \hline\hline
\multirow{2}{*}{Music} & Hit@10 & 15.02$\pm$0.27 & 15.45$\pm$0.31 & \textbf{15.92$\pm$0.21} \\
 & NDCG@10 & 8.02$\pm$0.15 & 8.28$\pm$0.22 & \textbf{8.56$\pm$0.13} \\ \cline{2-5} 
\multirow{2}{*}{Movie} & Hit@10 & 14.97$\pm$0.23 & 16.01$\pm$0.33 & \textbf{16.33$\pm$0.28} \\
 & NDCG@10 & 6.93$\pm$0.20 & 8.02$\pm$0.20 & \textbf{8.39$\pm$0.17} \\ \hline\hline
\multirow{2}{*}{Phone} & Hit@10 & 17.93$\pm$0.22 & 18.39$\pm$0.25 & \textbf{18.74$\pm$0.31} \\
 & NDCG@10 & 9.56$\pm$0.18 & 10.01$\pm$0.16 & \textbf{10.19$\pm$0.22} \\ \cline{2-5} 
\multirow{2}{*}{Elec} & Hit@10 & 20.03$\pm$0.25 & 20.37$\pm$0.28 & \textbf{21.03$\pm$0.29} \\
 & NDCG@10 & 10.97$\pm$0.16 & 11.21$\pm$0.24 & \textbf{11.54$\pm$0.23} \\ \hline\hline
\end{tabular}%
}
\end{table}
We further evaluate \method{} by examining the effectiveness of the momentum updating strategy and the stop-gradient operation of the target encoder. 
To assess their impact, we compare \method{} with two variants:
(1) Variant 1: It directly uses the online encoder as the target encoder and allows backpropagation through the target encoder.
(2) Variant 2: Similar to Variant 1, this variant uses the online encoder as the target encoder but stops the gradient flow.
As shown in Table \ref{tab:siamese}, the performance of \method{} is significantly influenced by both the momentum updating strategy and the stop-gradient operation. The momentum updating strategy ensures that the target encoder evolves more smoothly, capturing stable and consistent representations. 
Meanwhile, the stop-gradient operation prevents the online and target encoders from collapsing into identical representations, thereby preserving the model’s ability to learn diverse and disentangled features. These mechanisms play a vital role in maintaining the effectiveness of the contrastive learning framework and ensuring robust cross-domain recommendation performance.

\section{Derivation of ELBO}
Here we show how to obtain the ELBO in Equation 13. 
Given user $u_i$, its surrogate label $u_j$ and the index $k$ of its latent intents pertinent to both domains, $\log p(u_j|u_i)$ can be written as:
\begin{equation*}
\small
\begin{split}
    &\log p(u_j|u_i) \\
    &= \log \frac{p(u_i, u_j)p(k|u_i, u_j)}{p(u_i)p(k|u_i,u_j)}\\
    &=\log \frac{p(u_i, u_j, k)}{p(u_i)p(k|u_i, u_j)}\\
    &=\log \frac{p(u_j|u_i, k)p(u_i, k)}{p(u_i)p(k|u_i, u_j)}\\
    &=\log \frac{p(u_j|u_i, k)p(k|u_i)}{p(k|u_i,u_j)}\\
    &=\log p(u_j|u_i, k)-\log p(k|u_i, u_j)+\log p(k|u_i)\\
    &=\log p(u_j|u_i, k)-\log\frac{p(k|u_i, u_j)}{q(k|u_i, u_j)}+\log\frac{p(k|u_i)}{q(k|u_i, u_j)}.
\end{split}
\end{equation*}
We integrate both sides of the above equation. 
For the left-hand side,
\begin{equation*}
\small
\begin{split}
    \int q(k|u_i,u_j)\log p(u_j|u_i)dk&=\log p(u_j|u_i)\int q(k|u_i,u_j)dk\\
    &=\log p(u_j|u_i).
\end{split}
\end{equation*}
For the right-hand side,
\begin{equation*}
\small
\begin{split}
    \!&\int\!q(k|u_i,u_j)\log p(u_j|u_i, k)dk\!-\!\int\!q(k|u_i,u_j)\log\frac{p(k|u_i, u_j)}{q(k|u_i, u_j)}dk\\
    \!&\!+\!\int\!q(k|u_i,u_j)\log\frac{p(k|u_i)}{q(k|u_i, u_j)}dk\\
    \!&\!=\!\mathbb{E}_{q(k|u_i,u_j)}\log p(u_j|u_i,k)\!+\!D_{KL}(q(k|u_i,u_j) \Vert p(k|u_i,u_j))\\
    \!&\!-\!D_{KL}(q(k|u_i,u_j) \Vert p(k|u_i))\\
    \!&\geq \mathbb{E}_{q(k|u_i,u_j)}\log p(u_j|u_i,k)\!-\!D_{KL}(q(k|u_i,u_j) \Vert p(k|u_i)).
\end{split}
\end{equation*}
The equality holds when \begin{small}$D_{KL}(q(k|u_i,u_j) \Vert p(k|u_i,u_j))\geq 0$\end{small}. 
Consequently, this foundational premise facilitates the derivation of the Evidence Lower Bound (ELBO).

\bibliography{aaai25}